\documentclass[12pt]{iopart}
\usepackage{iopams}  
\expandafter\let\csname equation*\endcsname=\relax
\expandafter\let\csname endequation*\endcsname=\relaxD
\usepackage{epsf,graphicx,graphics}
\usepackage{amsmath}
\usepackage{latexsym,amssymb}
\usepackage{mathtools}
\usepackage{setspace,cite}
\usepackage{a4}
\usepackage{slashed}
\usepackage{float}
\usepackage[normal]{caption}

\usepackage[left=3cm,right=3cm,top=3cm,bottom=4cm]{geometry}

\newcommand{\od}[2]{\frac{d #1}{d #2}}

\usepackage{panmaths}

\begin{document}

\renewcommand{\thefootnote}{\arabic{footnote}}
\title[Furry dyons with $\Lambda<0$]{Globally regular solutions to dyonic anti-de Sitter $\mk{su}(\infty)$ Einstein-Yang-Mills theory -- Existence and characterising charges}
\author{J. Erik Baxter}
\address{Physics Research Group,
	Sheffield Hallam University,
	1 Howard Street,
	Sheffield, 
	South Yorkshire S1 1WB}
\ead{e.baxter@shu.ac.uk}
\begin{abstract}
	In this work, we find new static, spherically symmetric, dyonic, \emph{globally regular} exact solutions to $\mk{su}(\infty)$ Einstein-Yang-Mills theory with a negative cosmological constant $\Lambda$, in the regime that $|\Lambda|$ is very large. In this regime, we also prove that dyonic globally regular solutions may be uniquely characterised by a countably infinite set of effective global charges; and that dyon solutions may be distinguished from dyonic black hole solutions to the same field equations by their ADM masses. These solutions have potential modelling applications for certain exotic gravitational objects.
\end{abstract}
%

\maketitle
\vspace{1cm}

\section{Introduction}\label{Intro}

Einstein-Yang-Mills (EYM) models with large diffeomorphism groups, both classical and quantum-theoretical, are a subject of much current attention. Indeed, Hawking's final work concerned black hole models in string-theoretical versions of these systems, concerning the so-called \emph{Virasoro algebra} \cite{hawking_soft_2016,haco_black_2018}. String theories are characterised by infinite-rank Lie groups, because the infinitude of degrees of freedom can be used to represent the infinite possible string states. The infinite-rank Virasoro algebra is isomorphic to the algebra $w_\infty$, the group of area-preserving diffeomorphisms on the string world-sheet, which is in turn isomorphic to the group of area-preserving diffeomorphisms on that world-sheet surface SDiff$(\Sigma^2)$ for some 2D surface $\Sigma^2$. In the case where $\Sigma^2=S^2$, the 2-sphere, then these latter two algebras are also isomorphic to the algebra $\suinf$ \cite{bakas_large_1989,sezgin_area-preserving_1992,pope_higher-spin_1990}. This coincidence was noticed in \cite{hoppe_quantum_1982} and expanded by \cite{floratos_note_1989} into a dictionary of correspondences which can consistently define $\sun$ in the limit $N\rar\infty$. Our research concerns classical versions of these models.

Black hole solutions to models endowed with large diffeomorphism groups have therefore attracted a great deal of research interest (see e.g.  \cite{baxter_soliton_2007,cardoso_testing_2016,hawking_superrotation_2016,mirbabayi_dressed_2016,gabai_large_2016,baxter_existence_2018}), and this is for a few reasons. Such models have recently been used to investigate the Black Hole Information Paradox \cite{hawking_soft_2016,haco_black_2018,ellis_w_infty_2016,antoniadis_proceedings_2015}, since it is argued that infinite-dimensional Lie algebras contain the requisite `space' to contain the very large number of degrees of freedom which accompanies the Bekenstein-Hawking entropy \cite{bekenstein_statistical_1975}, and hence possibly can `encode' the information in some way \cite{hawking_black_1976}. Another reason they have attracted interest, in the case of a negative cosmological constant $\Lambda$, is due to the Anti-de Sitter/Conformal Field Theory (AdS/CFT) correspondence \cite{maldacena_large_1998}. The presence of black hole hair corresponds to observables in the dual CFT \cite{aharony_large_2000,hertog_black_2004}. The limit as the rank of the gauge group goes to infinity is the limit in which AdS/CFT becomes `exact' in the sense that it will approximate supergravity on the D-brane \cite{nastase_introduction_2015}. \emph{Dyonic} black holes, i.e. solutions in models which include a non-trivial electric sector of the gauge field (see e.g. \cite{baxter_existence_2016,baxter_existence_2018,nolan_stability_2016}), have been used via this correspondence to model holographic superconductors \cite{hendi_holographical_2018,cai_introduction_2015,shepherd_black_2017}. Also, there is ever the question of further verifying or expanding Bizon's modified No-Hair conjecture \cite{bizon_gravitating_1994}, which attempts to classify stable black holes (in a theory whose Lagrangian contains matter terms) by asymptotically defined charges. This too is a subject of current research \cite{winstanley_menagerie_2015,baxter_abundant_2008,baxter_stability_2016,shepherd_characterizing_2012,baxter_inprep_stab}.

As well as black holes, much recent research has concerned \emph{solitons}, which are purely magnetic, or \emph{dyons} if they also involve the electric sector \cite{baxter_soliton_2007,ponglertsakul_solitons_2016,kichakova_su2_2017,perapechka_su2_2018,herdeiro_gravitating_2018,baxter_existence_2018,baxter_existence_2015,shepherd_characterizing_2012}. These are a less famous but nevertheless interesting class of objects which began their life as Bartnik and McKinnon's `particle-like solutions' to purely magnetic, asymptotically flat $\essu$ EYM field equations \cite{bartnik_particle-like_1988}. Mathematically, they are \emph{globally regular}, in that all field variables are non-singular throughout the spatial range. They therefore do not have an event horizon or singularity, and can be viewed as black holes in the limit that the event horizon radius $r_h\rar0$. The historical interest in them has been for a few reasons -- partly because of the more complicated but elegant analytical methods which are generally necessary in these cases, but also because of the possibility of modelling applications, which includes so-called \emph{strange stars}, such as (mini) `boson stars' \cite{lee_soliton_1987,friedberg_scalar_1987,herdeiro_gravitating_2018}. Boson stars, and other such objects, have recently attracted attention in research as a possible producer of gravitational waves during binary mergers \cite{palenzuela_gravitational_2017,bezares_gravitational_2018,croon_boson_2018}. In the view of the author, there may be the possibility that solitons and dyons in $\suinf$ EYM theory and in $\sun$ EYM theories for very large $N$ could be used to model \emph{black hole remnants}, one of the possible hypothesised end products of Hawking evaporation \cite{chen_black_2014}: Solitons are very small, low-mass objects with the large number of degrees of freedom that are associated with black hole remnants \cite{hawking_black_1976}. This may be worthy of investigation.

Recently, we derived field equations for the dyonic $\suinf$ AdS EYM field theory, and investigated black hole solutions to this system \cite{baxter_inprep}, which in turn followed an analysis of the purely magnetic system in \cite{mavromatos_infinitely_2000}. We were able to prove the existence of solutions in two separate regimes: Solutions where the gauge fields are sufficiently small; and solutions in the limit that $|\Lambda|\rar\infty$. In addition, we were able to establish the existence of effective charges that uniquely define the latter class of black hole solutions by their asymptotic properties. However, since globally regular solutions exist to $\sun$ EYM theory in both the purely magnetic and dyonic cases, it is of interest to ask whether the same is true of $\suinf$, and whether those solutions can also be uniquely specified by their asymptotic data. Furthermore, if we wish the charges we defined to truly characterise each solution, we must establish that we may in general distinguish globally regular solutions from black hole solutions to the same system. Hence, these are the goals of this work.

\section{Field equations for $\mk{su}(\infty)$ Einstein-Yang-Mills theory}\label{AnsFE}

In this Section, we set the scene by outlining the dyonic $\suinf$ field equations. These have been previously derived \cite{baxter_inprep}, so we state the details of the system here.

The $\sun$-invariant\footnote{For the rest of this work, reference to $\sun$ implies that $N$ is finite, in contrast to $\suinf$.} gauge potential for the system is well known \cite{baxter_existence_2016}, and may be written as:
\begin{equation}\label{gaugepot}
\mc{W}\equiv\mc{W}_\mu dx^\mu\equiv Adt+Bdr+\frac{1}{2}\left(C-C^\dagger\right)d\theta-\frac{i}{2}\left[\left(C+C^\dagger\right)\sin\theta+D\cos\theta\right]d\phi.
\end{equation}
In $\sun$, the objects $A$, $B$, $C$ and $D$ are $(N-1)\times(N-1)$ matrices which contain functions of the radial co-ordinate $r$. In the limit $N\rar\infty$, we find \cite{hoppe_quantum_1982,floratos_note_1989,mavromatos_infinitely_2000} that $A$, $B$, $C$ and $D$ become functions of the radial co-ordinate $r$ and two `internal' angular co-ordinates $(\vartheta,\varphi)$ which parametrise spherical harmonics. Thus we may expand them as follows:
\begin{equation}\label{ansatzsumY}
\begin{matrix}
A=\slim_{l=1}^\infty a_l(r)P^0_l(\cos\vartheta), & \quad & B=\slim_{l=1}^\infty b_l(r)P^0_l(\cos\vartheta),\\
& \\
C=\slim_{l=1}^\infty c_l(r)P^1_l(\cos\vartheta)e^{i\varphi}, & \quad & D=2P^0_1(\cos\vartheta),
\end{matrix}
\end{equation}
where $P^0_l(\cos\vartheta)$ and $P^1_l(\cos\vartheta)$ are Legendre functions of orders 0 and 1 respectively, defined using Rodrigues' formula:
\begin{equation}\label{PL}
P^0_l(\cos\vartheta)=\frac{1}{2^l l!}\left(\frac{d}{d(\cos\vartheta)}\right)^l\left(\cos^2\vartheta-1\right)^l,
\end{equation}
and
\begin{equation}\label{PL1}
P^1_l(\cos\vartheta)=\left(1-\cos^2\vartheta\right)^\frac{1}{2}\left(\frac{d}{d(\cos\vartheta)}\right)P^0_l(\cos\vartheta).
\end{equation}
We would expect the sums in \eqref{ansatzsumY} to converge, which for instance would imply that the functions $a_l$, $b_l$ and $c_l$ are uniformly bounded $\forall \,r,\,l$, and that the magnitudes of coefficient functions in \eqref{ansatzsumY} drop off sufficiently quickly for large $l$. This is necessary for the physicality of solutions. However, we will not assume this, and will return to this issue in Section \ref{converge}.

The field equations imply that we may make a particular choice of gauge, and set $B\equiv0$. Then, letting $\xi\equiv\cos\vartheta$, we may write out the other gauge potential functions more explicitly as
\begin{equation}\label{ansatzsumxi}
A(r,\xi,\varphi)=\frac{i}{2}\alpha(r,\xi),\quad C(r,\xi,\varphi)=\omega(r,\xi)e^{i\varphi},\quad D=-2\xi
\end{equation}
for real functions $\alpha$ and $\omega$. Examining \eqref{ansatzsumY}, \eqref{PL} and \eqref{PL1}, we also have the following useful forms for the gauge functions, expressed as power series in terms $\xi$, expanded about the point $\xi=0$:
\begin{equation}\label{Ansatze}
\begin{split}
\quad\alpha(r,\xi)&=\slim_{j=1}^\infty\alpha_j(r)\xi^{j},\\
\omega(r,\xi)&=\sqrt{1-\xi^2}\slim_{j=0}^\infty\omega_j(r)\xi^j.
\end{split}
\end{equation}
Finally, we note that since $A$ is an element in the Cartan subalgebra, then the function $\alpha$ must obey the constraint
\begin{equation}\label{alphaint}
\int_{-1}^1\alpha(r,\xi)d\xi=0.
\end{equation}

We are now ready to present the field equations. The derivation begins with the well-known asymptotically AdS EYM field equations for $\sun$ \cite{baxter_soliton_2007}. As $N\rar\infty$, they may be written in the form \cite{mavromatos_infinitely_2000}
\begin{equation}
G_{\mu\nu}+\Lambda g_{\mu\nu}=\,\kappa T_{\mu\nu},\qquad
\nabla_\lambda F^\lambda_{\,\,\mu}+i\{\mc{W}_\lambda,F^\lambda_{\,\,\mu}\}=\,0,
\end{equation}
where in SI units $\kappa=8\pi G/c^4$, but as in \cite{baxter_inprep}, we choose units where $\kappa=3$ so that the equations may be embedded in $\essu$, a natural requirement since $\essu$ is a proper subalgebra of $\sun\,\,\forall N>2$ and hence of $\suinf$. Here, $G_{\mu\nu}$ is the Einstein tensor, the cosmological constant is $\Lambda<0$, and as in \cite{mavromatos_infinitely_2000} we can write the energy-momentum tensor as
\begin{equation}\label{T}
T_{\mu\nu}=\frac{1}{2}\int^1_{-1}\left[2g^{\rho\sigma}F_{\rho\mu}F_{\sigma\nu}-\frac{1}{2}g_{\mu\nu}F_{\rho\sigma}F^{\rho\sigma}\right]d\xi.
\end{equation} 
In \eqref{T}, the antisymmetric field strength tensor $F_{\mu\nu}$ is defined in terms of the gauge potential as
\begin{equation}
F_{\mu\nu}=\partial_\mu\mc{W}_\nu-\partial_\nu\mc{W}_\mu+i\{\mc{W}_\mu,\mc{W}_\nu\}.
\end{equation}
In all of the above, we have defined the Poisson bracket $\{Q,R\}$:
\begin{equation}
\{Q,R\}=\frac{\partial Q}{\partial\xi}\frac{\partial R}{\partial\varphi}-\frac{\partial R}{\partial\xi}\frac{\partial Q}{\partial\varphi}.
\end{equation}
As for the metric, we use the signature $(-++\,+)$, and then the line element can be written as
\begin{equation}
ds^2=-\mu S^2dt^2+\mu^{-1}dr^2+r^2\left(d\theta^2+\sin^2\theta d\phi^2\right),
\end{equation}
along with standard spherically symmetric `Schwarzschild-type' co-ordinates $(t,r,\theta,\phi)$, and we emphasise that $(\theta,\phi)$ are distinct from the internal co-ordinates $(\vartheta,\varphi)$. Because we are interested in static solutions, we take $\mu$ and $S$ (the \emph{lapse} function) to be functions of $r$ alone. The \emph{mass fraction} $\mu$ may be expressed as
\begin{equation}\label{mu}
\mu(r)=1-\frac{2m(r)}{r}+\frac{r^2}{\ell^2},
\end{equation} 
in which $m(r)$ is known as the \emph{mass function}, representing the effective mass of the solution, and $\ell$ is the \emph{AdS radius of curvature}, defined as
\begin{equation}\label{ell}
\ell=\sqrt{\frac{-3}{\Lambda}}. 
\end{equation}
This metric reduces to the ordinary Schwarzschild-anti-de Sitter metric function when $m(r)$ is a constant. Note that we are interested in globally regular solutions, so that we are working in the range $r\in[0,\infty)$.

With these assumptions and definitions, the field equations are as follows. The Einstein equations are given by
\begin{equation}\label{EEs}
m^\prime=\frac{3}{2}\left(\frac{r^2\eta}{4S^2}+\frac{\zeta}{4\mu S^2}+\mu G+\frac{P}{r^2}\right),\qquad
\frac{S^\prime}{S}=\frac{2}{r}\left(G+\frac{\zeta}{4\mu^2 S^2}\right),
\end{equation}
and the Yang-Mills equations can be written
\begin{equation}\label{YMs}
\begin{split}
0&=r^2\mu\frac{\partial^2\alpha}{\partial r^2}+r^2\mu\left(\frac{2}{r}-\frac{S^\prime}{S}\right)\frac{\partial\alpha}{\partial r}+\frac{\partial}{\partial\xi}\left(\omega^2\frac{\partial\alpha}{\partial\xi}\right),\\
0&=r^2\mu\frac{\partial^2\omega}{\partial r^2}+\left(2m-\frac{3P}{r}+\frac{2r^3}{\ell^2}-\frac{3 r^3\eta}{4S^2}\right)\frac{\partial\omega}{\partial r}+\frac{r^2\omega}{4\mu S^2}\left(\frac{\partial\alpha}{\partial\xi}\right)^2\\
&\,\,\,\,\,+\frac{\omega}{2}\frac{\partial^2}{\partial\xi^2}\left(\omega^2+\xi^2\right),
\end{split}
\end{equation}
where we have defined the quantities
\begin{equation}\label{xiquants}
\begin{split}
\eta&=\frac{1}{2}\int_{-1}^1\left(\frac{\partial\alpha}{\partial r}\right)^2d\xi,\\
\zeta&=\frac{1}{2}\int_{-1}^1\omega^2\left(\frac{\partial\alpha}{\partial\xi}\right)^2d\xi,\\
G&=\frac{1}{2}\int_{-1}^1\left(\frac{\partial\omega}{\partial r}\right)^2d\xi,\\
P&=\frac{1}{8}\int_{-1}^1\left(\frac{\partial}{\partial\xi}\left(\omega^2+\xi^2\right)\right)^2d\xi.
\end{split}
\end{equation}
Note that each of the expressions in \eqref{xiquants} are positive, and therefore from \eqref{EEs} both $m(r)$ and $S(r)$ will be monotonic functions.
%

\section{Boundary conditions and symmetries}\label{BCs}

Since we are considering globally regular solutions, we consider the whole spatial range for $r$. Therefore, solutions to will be defined on the semi-infinite strip $(r,\xi)\in\mc{R}\times\mc{I}$, where we define
\begin{equation}\label{RI}
\mc{R}\equiv[0,\infty),\qquad\mbox{ and }\qquad\mc{I}\equiv[-1,1].
\end{equation}
We also note that the field equations remain invariant under the three independent symmetries 
\begin{equation}\label{mirsym}
\xi\mapsto-\xi,\quad \alpha(r,\xi)\mapsto-\alpha(r,\xi), \quad\omega(r,\xi)\mapsto-\omega(r,\xi),
\end{equation}
and also the scaling symmetry
\begin{equation}\label{scalesym}
t\mapsto\tau^{-1} t,\quad S\mapsto\tau S,\quad \alpha\mapsto\tau\alpha,
\end{equation}
for some $\tau\in\R$. 

We will now briefly review the relevant boundary conditions for solutions to the field equations.

\subsection{Boundary conditions for $\xi=\pm1$}

These are inherited from the boundary conditions of Legendre functions. Examining the form of \eqref{Ansatze}, we can see that the boundary conditions for $\omega$ on the $\xi$-boundaries are
\begin{equation}\label{omegaxibcs}
\omega(r,\pm1)\equiv 0\,\,\,\forall r\in\mc{R}.
\end{equation}
The $\xi$-boundary conditions for the electric field $\alpha$ are less clear, since $P^0_l(\pm 1)=(\pm 1)^l$. It appears as though they must be calculated once the solutions are known, upon considering \eqref{ansatzsumY}. However, the regularity of solutions at the $\xi$ boundaries will follow naturally from the global regularity of the solutions themselves, which we address in Section \ref{LargeLambda}.

\subsection{Boundary conditions at $r=0$}\label{BC0}

Here, we expand all field variables in power series about the point $r=0$ in the form 
\begin{equation}
f=f_0+\slim_{j=1}^\infty f_jr^j, 
\end{equation}
where $f_0$ and $\{f_j\}$ are constants for $m$ and $S$, and functions of $\xi$ for $\alpha$ and $\omega$, and we substitute them into the field equations (\ref{EEs}, \ref{YMs}). We start by examining the lower-order terms, and if we require that the field equations and the functions $\mu$ and $\mu^\prime$ are all regular at $r=0$, we obtain the following expansions:
\begin{equation}\label{solexp}
\begin{split}
m(r)&=\slim_{j=3}^\infty m_jr^j,\\
S(r)&=S_0+\slim_{j=2}^\infty S_jr^j,\\
\alpha(r,\xi)&=\slim_{j=1}^\infty\alpha_j(\xi)r^j,\\
\omega(r,\xi)&=\sqrt{1-\xi^2}+\slim_{j=2}^\infty\omega_j(\xi)r^j,
\end{split}
\end{equation}
for two infinite sets of constants $\{m_j,S_j\}$ and two infinite sets of functions $\{\alpha_j(\xi),\omega_j(\xi)\}$. What we wish to do now is to prove that the set $\{m_j,S_j,\alpha_j,\omega_j\}$ may be recursively calculated to arbitrarily high order.

First we will examine the gravitational sector \eqref{EEs}. Note that the mass function has the condition $m(0)=0$, as expected, and if we scale $S$ using \eqref{scalesym} so that $\lim_{r\rar\infty}S(r)=1$, for asymptotic flatness, then that fixes the value of $S_0$ which we note must be non-zero for regularity. As for the higher order terms, we may derive recursion relations for $m_j$ and $S_j$. To simplify the calculation we introduce the expansions
\begin{equation}
\mu S^2\equiv S_0^2+\slim_{j=2}^\infty\widehat{M}_jr^j,\qquad\mu^2 S\equiv S_0+\slim_{j=2}^\infty\widetilde{M}_jr^j,
\end{equation}
so that
\begin{equation}\label{Ms}
\begin{split}
\widehat{M}_j=&\frac{S_0^2}{\ell^2}\delta^2_j+2S_0\left(S_j-S_0m_{j+1}+\frac{1}{\ell^2}S_{j-2}\right)+\slim_{k=2}^{j-2}\left(S_kS_{j-k}-4S_0S_km_{j-k+1}\right)\\
&+\frac{1}{\ell^2}\slim_{k=2}^{j-2}S_kS_{j-k-2}-2\slim_{k=5}^{j-3}\slim_{m=3}^{k-2}m_{m}S_{k-m}S_{j-k+1},\\
\widetilde{M}_j=&\frac{S_0}{\ell^2}\left(2\delta^2_j+\frac{1}{\ell^2}\delta^4_j\right)-4S_0m_{j+1}-\frac{4S_0}{\ell^2}m_{j-1}+S_j+\frac{2}{\ell^2}S_{j-2}+\frac{1}{\ell^4}S_{j-4}\\
&+4S_0\slim_{k=2}^{j-2}m_{k+1}m_{j-k+1}-4\slim_{k=2}^{j-2}S_km_{j-k+1}-\frac{4}{\ell^2}\slim_{k=2}^{j-4}S_jm_{j-k+1}\\
&+4\slim_{k=4}^{j-2}\slim_{m=2}^{k-2}S_mm_{k-m+1}m_{j-k+1},\\
\end{split}
\end{equation}
where $\delta^k_j$ is the Kronecker delta. Note that for each $j$, $\widehat{M}_j$ and $\widetilde{M}_j$ depend on the coefficients $\{m_{k+1},S_{k}\}$ for $k\leq j$. We thus obtain the following recurrence relations for $S_{j+1}$ and $m_{j+1}$:
\begin{equation}\label{EESr0}
\begin{split}
jS_0S_{j}=\,& 2S_0^2G_{j}+\frac{\zeta_j}{2}-\slim_{k=2}^{j-2}\widetilde{M}_k(j-k)S_{j-k}+2S_0\slim_{k=2}^{j-2}G_{k}\left(\widetilde{M}_{j-k}+S_{j-k}\right)\\
&+\slim_{k=4}^{j-2}\slim_{m=2}^kG_m\widetilde{M}_{k-m}S_{j-k}.\\
\end{split}
\end{equation}
\begin{equation}\label{EEmr0}
\begin{split}
\frac{2S_0^2(j+1)}{3}m_{j+1}=\,&\frac{\eta_{j-2}}{4}+\frac{\eta_{j-4}}{4\ell^2}+S_0^2P_{j+2}+\frac{\zeta_j}{4}+S_0^2\left(G_j+\frac{G_{j-2}}{\ell^2}\right)\\
&-2S_0^2\slim_{k=3}^{j-1}m_{k}G_{j-k+1}+\slim_{k=2}^{j-2}\widehat{M}_kG_{j-k}+\frac{1}{\ell^2}\slim_{k=2}^{j-4}\widehat{M}_kG_{j-k-2}\\
&-\frac{2}{3}\slim_{k=2}^{j-2}(k+1)m_{k+1}\widehat{M}_{j-k}+\slim_{k=2}^{j-2}\widehat{M}_kP_{j-k+2}\\
&-\frac{1}{2}\slim_{k=0}^{j-4}\eta_{k}m_{j-k-1}-2\slim_{k=5}^{j-1}\slim_{m=2}^{k-3}\widehat{M}_mm_{k-m}G_{j-k+1},\\
\end{split}
\end{equation}
where
\begin{equation}\label{Equantsr0}
\begin{split}
\eta_j=&\frac{1}{2}\sum_{k=0}^{l}(k+1)(j-k+1)\int_{-1}^1\alpha_{k+1}\alpha_{j-k+1}d\xi,\\
\zeta_j=&\frac{1}{2}\int_{-1}^1\left[\slim_{k=2}^{j-2}(1-\xi^2)\od{\alpha_k}{\xi}\od{\alpha_{j-k}}{\xi}+2\sqrt{1-\xi^2}\slim_{k=3}^{j-1}\slim_{m=2}^{k-1}\omega_{m}\od{\alpha_{k-m}}{\xi}\od{\alpha_{j-k}}{\xi}\right.\\
&\left.+\slim_{k=5}^{j-1}\slim_{m=4}^{k-1}\slim_{p=2}^{m-2}\omega_p\omega_{m-p}\od{\alpha_{k-m}}{\xi}\od{\alpha_{j-k}}{\xi}\right]d\xi,\\
G_j=&\frac{1}{2}\slim_{k=0}^{j-2}(k+1)(j-k-1)\int_{-1}^1\omega_{k+1}\omega_{j-k-1}d\xi,\\
P_j=&\frac{1}{2}\int_{-1}^1\left[\slim_{k=2}^{j-2}\od{}{\xi}\left(\sqrt{1-\xi^2}\omega_k\right)\od{}{\xi}\left(\sqrt{1-\xi^2}\omega_{j-k}\right)\right.\\
&\left.+\slim_{k=4}^{j-2}\slim_{m=2}^{k-2}\od{}{\xi}\left(\sqrt{1-\xi^2}\omega_m\right)\od{}{\xi}\left(\omega_{k-m}\omega_{j-k}\right)\right.\\
&\left.+\frac{1}{4}\slim_{k=6}^{j-2}\slim_{m=4}^{k-2}\slim_{p=2}^{m-2}\od{}{\xi}\left(\omega_{p}\omega_{m-p}\right)\od{}{\xi}\left(\omega_{k-m}\omega_{j-k}\right)\right]d\xi.\\
\end{split}
\end{equation}

If we examine of each of the terms in (\ref{EESr0}, \ref{EEmr0}), and using \eqref{Ms} and \eqref{Equantsr0}, it can be seen that for each value of $j$, the coefficients $m_j$ depend only on $\{m_{k-1},S_{k-1},\alpha_{k-2},\omega_{k-1}\}$, and $S_j$ depend only on $\{m_{k},S_{k-1},\alpha_{k-2},\omega_{k-1}\}$, for all $k\leq j$.

The gauge sector is a lot more complicated, but it may be simplified a lot more than the metric sector. We can successively calculate the first three terms of each:
\begin{equation}\label{LOgauge}
\begin{split}
&\alpha_j(\xi)=\,\rho_jP^0_j(\xi)\,\,\,\mbox{ for }j=1,2,\quad\alpha_3(\xi)=\rho_3P^0_3(\xi)+\frac{\rho_1\sigma_2}{5}P^0_1(\xi);\\[10pt]
&\omega_j(\xi)=\,\sigma_jP^1_{j-1}(\xi)\,\,\,\mbox{ for }j=2,3,\\[10pt]
&\omega_4(\xi)=\sigma_4P^1_3(\xi)+\left(\frac{\rho_1^2}{40S_0^2}-\frac{3\sigma_2}{5\ell^2}+\frac{4\sigma_2^3}{3}-\frac{\rho_1^2\sigma_2}{12S_0^2}\right)P^1_1(\xi).
\end{split}
\end{equation}		
To calculate these we have used the lowest order coefficients of the metric sector, which are
\begin{equation}\label{LOmetric}
\begin{split}
& m_3=\dfrac{\rho_1^2}{8S_0^2}+2\sigma_2^2;\qquad
S_2=\dfrac{\rho_1^2}{6S_0}+\dfrac{8S_0}{3}\sigma_2^2.
\end{split}
\end{equation} 
Above, $\rho_j$ and $\sigma_j$ are the arbitrary constants from \eqref{GFr0}.

The higher order terms must be calculated recursively, for which we will need to invoke induction. This is therefore a good point to introduce the following
\begin{prop}
	The constants $\{m_j,S_j\}$ and the functions $\{\alpha_j(\xi),\omega_j(\xi)\}$ in \eqref{solexp}, for any fixed $j$, may be calculated from $\{m_k,S_k,\alpha_k(\xi),\omega_k(\xi)\}$ with $k<j$; and therefore the expansion terms in \eqref{solexp} may be calculated to arbitrarily high order. Furthermore, the gauge sector may be explicitly calculated: The gauge equations \eqref{YMs} become
	\begin{equation}\label{solGFprop}
	\begin{split}
	j(j+1)\alpha_{j}+\od{}{\xi}\left((1-\xi^2)\od{\alpha_{j}}{\xi}\right)&=\slim^{j-2}_{m=1}\beta^j_mP^0_m,\\
	j(j-1)\omega_{j}+\sqrt{1-\xi^2}\od{^2}{\xi^2}\left(\sqrt{1-\xi^2}\omega_{j}\right)&=\slim^{j-3}_{m=1}\gamma^j_mP^1_m,
	\end{split}
	\end{equation}
	for all $j$. For each $j$, $\beta^j_m$ and $\gamma^j_m$ depend only on the terms $\{m_{m},S_{m},\alpha_{m-1},\omega_{m}\}$ for all $m<j$; and \eqref{solGFprop} can be solved to yield
	\begin{equation}\label{GFr0indprop}
	\begin{split}
	\alpha_j(\xi)&=\rho_jP^0_j(\xi)+\slim_{m=1}^{j-2}\frac{\beta^j_{m}P^0_{m}(\xi)}{j(j+1)-m(m+1)},\\
	\omega_j(\xi)&=\sigma_jP^1_{j-1}(\xi)+\slim_{m=1}^{j-3}\frac{\gamma^j_{m}P^1_{m}(\xi)}{j(j-1)-m(m+1)}.
	\end{split}
	\end{equation}
\end{prop}

\textbf{Proof} As we recursively calculate the gauge field expansion coefficients, we see that each order of the gauge field may be written as a finite sum over Legendre functions, including one new arbitrary constant at every order. Therefore, we make the inductive hypothesis that we possess all the expansion coefficients $\{m_k,S_k,\alpha_k(\xi),\omega_k(\xi)\}$ with $k<J$, for $J$ some integer, and in addition, that $\alpha_k(\xi)$ and $\omega_k(\xi)$ conform to \eqref{GFr0indprop} for all $k<J$.

In that case, we write out the gauge field equations \eqref{YMs} at expansion order $J$. These may be written in the following form:
\begin{equation}\label{solGF}
\begin{split}
J(J+1)\alpha_{J}+\od{}{\xi}\left((1-\xi^2)\od{\alpha_{J}}{\xi}\right)&=\slim^{J-2}_{k=1}\beta^J_kP^0_k,\\
J(J-1)\omega_{J}+\sqrt{1-\xi^2}\od{^2}{\xi^2}\left(\sqrt{1-\xi^2}\omega_{J}\right)&=\slim^{J-3}_{k=1}\gamma^J_kP^1_k,
\end{split}
\end{equation}
where we have written the right-hand sides uniquely as a finite sum over Legendre functions $P^0_k(\xi)$ and $P^1_k(\xi)$; this is possible because of the forms of the functions of $\xi$ that appear on the right-hand sides. The expansion constants $\beta^J_k$ and $\gamma^J_k$ would take too long to write out in general: the point to make is that they depend only on the terms $\{m_{k},S_{k},\alpha_{k-1},\omega_{k}\}$ for all $k<J$. 

Equations \eqref{solGF} are inhomogeneous versions of the zeroeth and first order Legendre equations respectively, and since they are linear ODEs, we may employ the usual `complimentary function/particular integral' method. Their solutions can thus be calculated as:
\begin{equation}\label{GFr0}
\begin{split}
\alpha_J(\xi)&=\rho_JP^0_J(\xi)+\slim_{k=1}^{J-2}\frac{\beta^J_{k}P^0_{k}(\xi)}{J(J+1)-k(k+1)},\\
\omega_J(\xi)&=\sigma_JP^1_{J-1}(\xi)+\slim_{k=1}^{J-3}\frac{\gamma^J_{k}P^1_{k}(\xi)}{J(J-1)-k(k+1)},
\end{split}
\end{equation}
where we have chosen two new arbitrary constants $\{\rho_J,\sigma_J\}$. This matches the form \eqref{GFr0indprop}.

By the inductive hypothesis, we possess expressions for the set of expansion coefficients $\Sigma_{J-1}\equiv\{m_{k},S_{k},\omega_{k},\alpha_{k}\,\,|\,\,\forall\,k\leq J-1\}$. We may first use the obtained coefficients $\Sigma_{J-1}$, along with (\ref{Ms} - \ref{Equantsr0}), to calculate $m_{J}$. Then we may use $\{m_{J}\}\,\cup\,\Sigma_{J-1}$ to calculate $\beta^J_{k}$ for all $k\leq J-2$ and $\gamma^J_{k}$ for all $k\leq J-3$. This allows us to calculate $\{\alpha_{J},\omega_{J}\}$ \eqref{GFr0}. Finally, we may substitute the coefficients $\{m_{J},\alpha_{J},\omega_{J}\}\,\cup\,\Sigma_{J-1}$ into \eqref{EESr0} to calculate $S_{J}$. Thus, we obtain the set of expansion coefficients $\Sigma_{J}=\{m_{J},S_{J},\omega_{J},\alpha_{J}\}\cup\Sigma_{J-1}$; and so it is clear that by induction we may calculate the expansion coefficients in \eqref{solexp} to arbitrarily high order, and the Proposition is proven. $\Box$

We note that when the expressions \eqref{GFr0} are calculated and substituted back into \eqref{solexp}, then $\alpha$ and $\omega$ conform to the ansatz in the form \eqref{ansatzsumY}. We further note that one new constant of integration is needed for every order of $r$ in the series, something which mirrors the case of $\sun$ \cite{baxter_existence_2008, baxter_existence_2016}. However, unlike the $\sun$ case, the infinitude of gauge degrees of freedom means that the boundary conditions here are \emph{exact}: No `higher order terms' need to be (or can be) included. Therefore, the solution is uniquely described near $r=0$ by an infinite set of constants, as we may expect.

\subsection{Boundary conditions as $r\rar\infty$}\label{BCinf}

The asymptotic boundary conditions are much simpler. These exactly the same in general as in the case of black holes \cite{baxter_inprep}, so we merely need to quote the results here. Changing variables to $z=r^{-1}$, and expressing the field variables in power series good near $z=0$, we find that the field equations \eqref{EEs}, \eqref{YMs} necessitate the following asymptotic behaviour:
\begin{equation}\label{infexp}
\begin{split}
m(z)&=M+m_1z+O(z^2),\\
S(z)&=1+O(z^4),\\
\alpha(z,\xi)&=\alpha_\infty(\xi)+\mc{A}(\xi)z+O(z^2),\\
\omega(z,\xi)&=\omega_\infty(\xi)+\mc{W}(\xi)z+O(z^2).\\
\end{split}
\end{equation}
Asymptotically, the mass function again has
\begin{equation}\label{m1}
\begin{split}
m_1=&\,-\frac{3}{2}\left(\frac{1}{8}\int\limits_{-1}^1\mc{A}^2d\xi+\frac{1}{8}\int\limits_{-1}^1\left(\pd{}{\xi}\left(\omega_\infty^2+\xi^2\right)\right)^2d\xi+\frac{\ell^2}{8}\int\limits_{-1}^1\omega_\infty^2\left(\pd{\alpha_\infty}{\xi}\right)^2d\xi\right.\\
&\,\left.
+\frac{1}{2\ell^2}\int\limits_{-1}^1\mc{W}^2d\xi\right).\\
\end{split}
\end{equation}
Therefore, in the asymptotic regime, solutions are specified completely by four functions of $\xi$ -- $\omega_\infty(\xi)$, $\alpha_\infty(\xi)$, $\mc{W}(\xi)=-\lim_{r\rar\infty}r^2\pd{\omega}{r}$, $\mc{A}(\xi)=-\lim_{r\rar\infty}r^2\pd{\alpha}{r}$ -- and the constant $M$, which we identify as the ADM mass of the solution.

\section{Globally regular solutions as $|\Lambda|\rar\infty$ ($\ell\rar 0$)}\label{LargeLambda}

\subsection{Existence of solutions}\label{LL}

When we investigated black hole solutions to the $\suinf$ AdS EYM system, we were able to prove existence of solutions in the case where both gauge fields are of order $\epsilon<<1$ for general $\Lambda<0$, and also in the case where $|\Lambda|\rar\infty$. However, if we repeat the analysis in \cite{baxter_inprep} for dyons, we find that the mass function must still be given by $m(r)=\mc{M}-\frac{1}{r}+O(\epsilon^2)$ for $\mc{M}$ some constant of integration. This is clearly singular at $r=0$, and so we find that globally regular solutions cannot exist if the gauge fields are too small, and this result applies to purely magnetic solitons and to dyons alike. This is a surprising result, since we can think of no reason \emph{a posteriori} why this should be the case. Hence, we concentrate on finding solutions for which $|\Lambda|$ is sufficiently large. 

There are other good reasons to consider this regime. In numerical models for $\sun$, regions of the initial parameter space admitting solutions shrinks as $N$ grows, but for any $N$, grows without limit as $|\Lambda|\rar\infty$ \cite{baxter_soliton_2007}. In addition, it is necessary for the stability of solutions in that case, and the definition of uniquely characterising charges \cite{baxter_existence_2008,baxter_stability_2016,shepherd_characterizing_2012}. Phenomenologically speaking, the regime corresponds to large black holes which possess a stable Hartle-Hawking state, and this is of relevance to the question of information loss during Hawking evaporation \cite{hawking_thermodynamics_1983}. It is also the regime where the field theory approximation may be used in string-theoretical treatments, in the sense that string corrections become negligible in the bulk \cite{maldacena_large_1998,witten_anti-sitter_1998}.

Our strategy will be to assume power series expansions for the field variables, using the AdS radius of curvature $\ell$ as the expansion parameter. We assume that $\ell$ is very small, which corresponds to a very large value of $|\Lambda|$ \eqref{ell}. These will be asymptotic expansions, and so in the limit where all terms of the expansion are taken into consideration, our solutions will be exact, provided that the sums defined by \eqref{ansatzsumY} converge. In fact as we will see, the solution is not a genuine $\suinf$ solution \emph{unless} we consider all expansion terms in the infinite sum, so we will also give attention to the convergence properties of the solutions we find. 

In the case of dyons, we must be careful about how we take the limit $\ell\rar0$. This is because in the black hole case, we have a natural `scale' to work with: the ratio between the event horizon radius $r_h$, and $\ell$. We must here do without the event horizon radius. 
Therefore, as in previous cases \cite{baxter_existence_2008,baxter_existence_2016}, we rescale all dimensionful quantities thus:
\begin{equation}\label{elltrans}
r=\ell x,\qquad m(r)=\ell\hat{m}(x\ell).
\end{equation}
If we substitute these into the field equations (\ref{EEs}, \ref{YMs}) and then let $\ell=0$, we obtain the unique solution
\begin{equation}\label{ell0}
\hat{m}\equiv 0,\qquad S(r)\equiv 1,\qquad \omega(r,\xi)=\sqrt{1-\xi^2},\qquad\alpha(r,\xi)\equiv 0.
\end{equation}
However, as in the black hole case, we \emph{cannot} consider the case $\ell=0$ to be meaningful since the variables \eqref{elltrans} become inapplicable; therefore we again consider the case where $\ell$ is small i.e. in some neighbourhood of zero, and attempt to find asymptotic series expansions using $\ell$ as our expansion parameter, which will therefore be in some sufficiently small neighbourhood of the solution \eqref{ell0}.

We find that the gauge field variables are most usefully expanded in very similar terms to their expansions near $r=0$ \eqref{solexp}. Calculating some of the lowest order terms and requiring global regularity, the expansions must take the form 
\begin{equation}\label{dyonLLexp}
\begin{split}
\hat{m}(x)&=\slim_{k=2}^\infty\hat{m}_k(x)\ell^k,\qquad S(x)=1+\slim_{k=2}^\infty\hat{S}_k(x)\ell^k,\\
\alpha(x,\xi)&=\slim_{k=1}^\infty\tilde{\rho}_k(x\ell)P^0_k(\xi)(x\ell)^k,\\
\omega(x,\xi)&=\sqrt{1-\xi^2}+\slim_{k=2}^\infty\tilde{\sigma}_k(x\ell)P^1_{k-1}(\xi)(x\ell)^k.
\end{split}
\end{equation}
It can be seen that the electric field $\alpha$ is small in the sense that $\alpha\sim O(\ell)$, and it is worth pointing out that this was by necessity rather than assumption, because it is the boundary conditions of the system that imply that $\alpha=0$ if $\ell=0$ \eqref{ell0}.

We first observe that since we require that these solutions satisfy the boundary conditions derived in Section \ref{BCs}, we choose the boundary conditions for $\hat{S}_k$ and $\hat{m}_k$ to be 
\begin{equation}\label{mkSkBC}
\lim_{x\rar\infty}\hat{S}_k(x)=0,\qquad\hat{m}_k(0)=0,
\end{equation}
for all $k\geq 2$. We note also that $\mu=1+x^2+O(\ell^2)$.

\begin{center}
	\begin{figure}[h]
		\centering
		\captionsetup{justification=centering}
		\includegraphics[scale=.55]{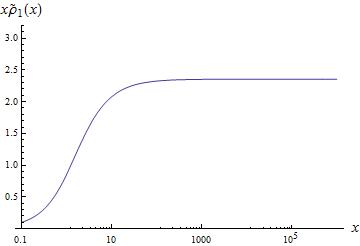}
		\caption{The function $x\tilde{\rho}_1(x)$, with $\rho_1=1$. This function tends to $\frac{3\pi}{4}$ as $x\rar\infty$.}\label{rho1}
	\end{figure}
\end{center}

\begin{center}
	\begin{figure}[h]
		\centering
		\captionsetup{justification=centering}
		\includegraphics[scale=.55]{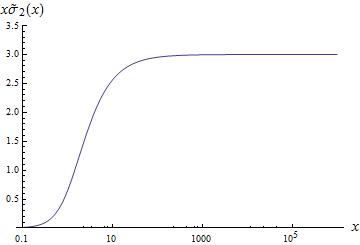}
		\caption{The function $x^2\tilde{\sigma}_2(x)$, with $\sigma_2=1$. This function tends to 3 as $x\rar\infty$.}\label{sig2}
	\end{figure}
\end{center}

We shall begin by calculating the lower order terms of the expansions, starting with the gauge sector. The equations for $\tilde{\rho}_1(x)$ and $\tilde{\sigma}_2(x)$ are homogeneous hypergeometric differential equations:
\begin{equation}
\begin{split}
x(1+x^2)\od{^2\tilde{\rho}_1}{x^2}+4(1+x^2)\od{\tilde{\rho}_1}{x}+2x\tilde{\rho}_1&=0,\\
x(1+x^2)\od{^2\tilde{\sigma}_2}{x^2}+2\left(2+3x^2\right)\od{\tilde{\sigma}_2}{x}+6x\tilde{\sigma}_2&=0,
\end{split}
\end{equation}
the solutions to which are given by
\begin{equation}
\begin{split}
\tilde{\rho}_1(x)&=\rho_1\left(_2F_1\left(\frac{1}{2}\,,\,1\,;\,\frac{5}{2}\,;\,-x^2\right)\right)=\frac{3\rho_1((1+x^2)\arctan x-x)}{2x^3},\\
\tilde{\sigma}_2(x)&=\sigma_2\left(_2F_1\left(1\,,\,\frac{3}{2}\,;\,\frac{5}{2}\,;\,-x^2\right)\right)=\frac{3\sigma_2(x-\arctan x)}{x^3},\\
\end{split}
\end{equation}
where $\rho_1=\tilde{\rho}_1(0)$ and $\sigma_2=\tilde{\sigma}_2(0)$ are constants which appear in \eqref{GFr0}. The functions $x\tilde{\rho}_1(x)$ and $x^2\tilde{\sigma}_2(x)$ govern the spatial behaviour, and these are plotted (letting $\rho_1=\sigma_2=1$) in Figures \ref{rho1} and \ref{sig2}. 

Knowing these, we may solve the lower order equations of the Einstein sector. The equations for $\hat{m}_2(x)$ and $\hat{S}_2(x)$ are
\begin{equation}\label{Ein2}
\begin{split}
\frac{d\hat{m}_2}{dx}&=\frac{x^2}{8}\left(\od{}{x}\left(x\tilde{\rho}_1\right)\right)^2+\frac{x^2\tilde{\rho}_1^2}{4(1+x^2)^2}+(1+x^2)\left(\frac{d}{dx}\left(x^2\tilde{\sigma}_2\right)\right)^2+2x^2\tilde{\sigma}_2^2,\\
\frac{d\hat{S}_2}{dx}&=\frac{4}{3x}\left(\frac{d}{dx}\left(x^2\tilde{\sigma}_2\right)\right)^2+\frac{x\tilde{\rho}_1^2}{3(1+x^2)^2}.
\end{split}
\end{equation}
We may use the properties of hypergeometric functions to see that \eqref{Ein2} implies the correct boundary behaviour for $\hat{m}_2$ and $\hat{S}_2$ as computed in Sections \ref{BC0} and \ref{BCinf}: Nearby $x=0$, we can see that $\od{\hat{m}_2}{x}\sim O(x^2)$ and $\od{\hat{S}_2}{x}\sim O(x)$; and as $x\rar\infty$, $\frac{d\hat{m}_2}{dx}\sim O\left(x^{-2}\right)$ and $\od{\hat{S}_2}{x}\sim O\left(x^{-5}\right)$. 

\begin{center}
	\begin{figure}[h]
		\centering
		\captionsetup{justification=centering}
		\includegraphics[scale=.55]{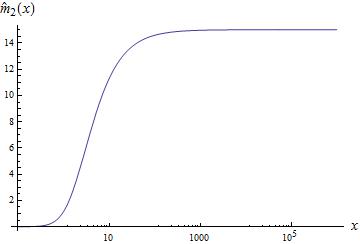}
		\caption{The function $\hat{m}_2(x)$, with $\rho_1=\sigma_2=1$.}\label{m2}
	\end{figure}
\end{center}

\begin{center}
	\begin{figure}[h]
		\centering
		\captionsetup{justification=centering}
		\includegraphics[scale=.55]{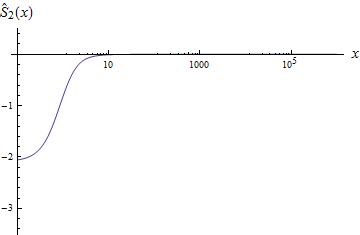}
		\caption{The function $\hat{S}_2(x)$, with $\rho_1=\sigma_2=1$.}\label{S2}
	\end{figure}
\end{center}

We seek numerical solutions solutions to \eqref{Ein2}, using the boundary conditions $\hat{m}_2(0)=0$ for the mass function and $\lim_{x\rar\infty}\hat{S}_2=0$ for the lapse function. Figures \ref{m2} and \ref{S2} show plots of the solutions to Equations  \eqref{Ein2}. We wish to note a few things about these solutions. Firstly, $\hat{m}_2(x)$ behaves as a physical mass function should, in that it is positive, monotonically increasing, and reaches a finite limit as $x\rar\infty$. Also, the truncated expansion $1+\hat{S}_2(x)\ell^2$ resembles $S(r)$ for a typical $\sun$ solution \cite{baxter_existence_2008}, in that it will start at some finite value, in this case just less than 1, and tend towards $1$ asymptotically.

Now we examine the higher order terms of the expansions \eqref{dyonLLexp}. The gauge equations \eqref{YMs} may be written as
\begin{equation}\label{hyperg}
\begin{split}
\left(x(1+x^2)\od{^2\tilde{\rho}_j}{x^2}+2(j+1)(1+x^2)\od{\tilde{\rho}_j}{x}+xj(j+1)\tilde{\rho}_k\right)P^0_j&=\slim_{m=1}^{j-2}\tilde{\beta}^j_m(x)P^0_m,\\[5pt]
\left(x(1+x^2)\od{^2\tilde{\sigma}_j}{x^2}+2\left(j+x^2(j+1)\right)\od{\tilde{\sigma}_j}{x}+xj(j+1)\tilde{\sigma}_j\right)P^1_{j-1}&=\slim_{m=1}^{j-3}\tilde{\gamma}^j_m(x)P^1_m,
\end{split}
\end{equation}
where the functions $\tilde{\beta}^j_m(x)$ and $\tilde{\gamma}^j_m(x)$ include sums of hypergeometric functions, their first derivatives, terms from the Einstein equations, and products of these. Happily, we can decouple the equations entirely by using orthogonality relations:
\begin{equation}
\int_{-1}^1P^0_j(\xi)P^0_k(\xi)d\xi=\frac{2}{2k+1}\delta_{j,k},\qquad\int_{-1}^1P^1_j(\xi)P^1_k(\xi)d\xi=\frac{2k(k+1)}{2k+1}\delta_{j,k},
\end{equation}
where $\delta_{j,k}$ is the Kronecker symbol. Hence, we multiply through by $P^0_k$ and $P^1_{k-1}$ respectively and integrate over the range $\mc{I}$. This yields the much simpler system
\begin{equation}\label{hyperghom}
\begin{split}
x(1+x^2)\od{^2\tilde{\rho}_k}{x^2}+2(k+1)(1+x^2)\od{\tilde{\rho}_k}{x}+xk(k+1)\tilde{\rho}_k&=0,\\[5pt]
x(1+x^2)\od{^2\tilde{\sigma}_k}{x^2}+2\left(k+x^2(k+1)\right)\od{\tilde{\sigma}_k}{x}+xk(k+1)\tilde{\sigma}_k&=0.
\end{split}
\end{equation}
It can be seen that with the transformation $z=-x^2$, the left hand sides of these equations are identical to those for hypergeometric differential equations, whose solutions can immediately be expressed as follows:
%
%
\begin{equation}\label{r0LLrho}
\tilde{\rho}_k(x)=\,\rho_k\mc{P}_k(x),\mbox{ where }\mc{P}_k\equiv\,_2F_1\left(\frac{k}{2}\,,\,\frac{k+1}{2}\,;\,\frac{2k+3}{2}\,;\,-x^2\right),
\end{equation}
and
\begin{equation}\label{r0LLsig}
\tilde{\sigma}_k(x)=\sigma_k\mc{S}_k(x),\mbox{ where }\mc{S}_k\equiv\,_2F_1\left(\frac{k}{2}\,,\,\frac{k+1}{2}\,;\,\frac{2k+1}{2}\,;\,-x^2\right),
\end{equation}
and where $\{\rho_k,\sigma_k\}$ are the constants defining the boundary conditions ar $r=0$ \eqref{GFr0}. The appearance of hypergeometric functions in the regime $|\Lambda|\rar\infty$ is something which is familiar from globally regular solutions in the $\sun$ case \cite{baxter_existence_2018,baxter_existence_2016,baxter_existence_2015}; but for $\suinf$ solutions, we get the added `bonus' that we can also calculate the gauge fields explicitly to all orders in $\ell$, which is a satisfying result.

When substituted back into \eqref{dyonLLexp}, then the $x$-dependence of the gauge field is entirely contained in the terms $x^k\tilde{\rho}_k$ and $x^k\tilde{\sigma}_k$. To examine the boundary behaviour, we may observe that $\mc{P}_k$ and $\mc{S}_k$ both behave like $1+O(x^2)$ near $x=0$, and asymptotically we use the identity \cite{abramowitz_handbook_1988} (15.3.7)
\begin{equation}\label{2F1asym}
\begin{split}
_2F_1\left(a,b;c;z\right)&=\frac{\Gamma(b-a)\Gamma(c)}{\Gamma(b)\Gamma(c-a)}(-z)^{-a}\,_2F_1\left(a,a-c+1;a-b+1;z^{-1}\right)\\
&+\frac{\Gamma(a-b)\Gamma(c)}{\Gamma(a)\Gamma(c-b)}(-z)^{-b}\,_2F_1\left(b,b-c+1;-a+b+1;z^{-1}\right).
\end{split}
\end{equation}
Then, we can compute the following behaviour at the spatial boundaries for $x^k\tilde{\rho}_k$ and $x^k\tilde{\sigma}_k$:
\begin{equation}\label{GFbounds}
\begin{split}
\lim_{x\rar 0}x^k\tilde{\rho}_k&=\lim_{x\rar 0}x^k\tilde{\sigma}_k=0,\\[5pt]
\lim_{x\rar\infty}x^k\tilde{\rho}_k&=\frac{\rho_k\sqrt{\pi}\,\Gamma\left(\frac{2k+3}{2}\right)}{\Gamma\left(\frac{k+1}{2}\right)\Gamma\left(\frac{k+3}{2}\right)},\quad\lim_{x\rar\infty}x^k\tilde{\sigma}_k=\frac{\sigma_k\sqrt{\pi}\,\Gamma\left(\frac{2k+1}{2}\right)}{\Gamma\left(\frac{k+1}{2}\right)^2}.
\end{split}
\end{equation}
Substituting these back into \eqref{dyonLLexp}, these imply the correct boundary conditions for $\alpha$ and $\omega$ \eqref{GFr0}.

Results for the metric sector are much more complicated, because the explicit expressions for the solutions $\hat{m}_j(x)$, $\hat{S}_j(x)$ are hard to find. Similar to Section \ref{BCs}, we make life easier by defining
\begin{equation}\label{widehat}
\mu S^2\equiv 1+x^2+\slim_{j=2}^{\infty}\widehat{\mc{M}}_j(x)\ell^j,\qquad \mu^2S\equiv (1+x^2)^2+\slim_{j=2}^{\infty}\widetilde{\mc{M}}_j(x)\ell^j,
\end{equation}
so that
\begin{equation}\label{RRM}
\begin{split}
\widehat{\mc{M}}_j&=2(1+x^2)\hat{S}_j-\frac{2}{x}\left(\hat{m}_j+2\slim_{k=2}^{j-2}\hat{m}_k\hat{S}_{j-k}\right)+(1+x^2)\slim_{k=2}^{j-2}\hat{S}_k\hat{S}_{j-k}\\
&-\frac{2}{x}\slim_{k=4}^{j-2}\slim_{m=2}^{k-2}\hat{m}_{m}\hat{S}_{k-m}\hat{S}_{j-k},\\
\widetilde{\mc{M}}_j&=(1+x^2)^2\hat{S}_j-\frac{4(1+x^2)}{x}\left(\hat{m}_j+\slim_{k=2}^{j-2}\hat{m}_k\hat{S}_{j-k}\right)\\
&+\frac{4}{x^2}\left(\slim_{k=2}^{j-2}\hat{m}_k\hat{m}_{j-k}+\slim_{k=2}^{j-4}\slim_{m=2}^{k}\hat{m}_m\hat{m}_{k-m}\hat{S}_{j-k}\right).
\end{split}
\end{equation}
Upon substituting \eqref{dyonLLexp}, \eqref{r0LLrho} and \eqref{r0LLsig} into the Einstein equations \eqref{EEs}, we obtain the following recursive differential equations:
\begin{equation}\label{RRm}
\begin{split}
\frac{2}{3}\od{\hat{m}_j}{x}&=-\frac{2}{3(1+x^2)}\slim_{k=2}^{j-2}\widehat{\mc{M}}_k\od{\hat{m}_{j-k}}{x}-\frac{x}{2(1+x^2)}\slim_{k=2}^{j-2}\widehat{\mc{M}}_k\hat{\eta}_{j-k-2}+\frac{\hat{\zeta}_j}{4(1+x^2)}\\
&+\slim_{k=2}^{j-2}\widehat{\mc{M}}_{k}\hat{G}_{j-k}-\frac{2}{x}\slim_{k=2}^{j-2}\hat{m}_k\hat{G}_{j-k}-\frac{2}{x(1+x^2)}\slim_{k=2}^{j-4}\slim_{m=2}^{k}\hat{m}_m\widehat{\mc{M}}_{k-m}\hat{G}_{j-k}\\
&+(1+x^2)\hat{G}_j+\frac{x^2}{4}\hat{\eta}_{j-2}+\frac{1}{x^2}\hat{P}_{j+2}+\frac{1}{x^2(1+x^2)}\slim_{k=4}^j\widehat{\mc{M}}_k\hat{P}_{j-k+2},
\end{split}
\end{equation}
\begin{equation}\label{RRS}
\begin{split}
\od{\hat{S}_j}{x}&=\frac{2}{x}\hat{G}_j+\frac{\hat{\zeta}_j}{2x(1+x^2)^2}+\frac{2}{x(1+x^2)^2}\slim_{k=2}^{j-2}\hat{G}_k\widetilde{\mc{M}}_{j-k}+\frac{2}{x}\slim_{k=2}^{j-2}\hat{G}_{k}\hat{S}_{j-k}\\
&+\frac{2}{x(1+x^2)^2}\slim_{k=4}^{j-2}\slim_{m=2}^{k-2}\hat{G}_m\hat{S}_{k-m}\widetilde{\mc{M}}_{j-k}-\frac{1}{(1+x^2)^2}\slim_{k=2}^{j-2}\widetilde{\mc{M}}_k\od{\hat{S}_{j-k}}{x},\\
\end{split}
\end{equation}
where we have expanded the quantities $\eta$, $G$, $\zeta$ and $P$ as 
\begin{equation}
\eta=\sum_{j=0}^\infty\hat{\eta}_j(x)\ell^j,\quad G=\sum_{j=2}^\infty \hat{G}_j(x)\ell^j,\quad\zeta=\sum_{j=2}^\infty\hat{\zeta}_j(x)\ell^j,\quad P=\sum_{j=4}^\infty \hat{P}_j(x)\ell^j,
\end{equation}
such that
\begin{equation}\nonumber
\begin{split}
\hat{G}_j&=\left\{\begin{matrix} \dfrac{j(j+2)}{4(j+1)}\left(\dfrac{d}{dx}\left(\tilde{\sigma}_{\frac{j}{2}+1}x^{\frac{j}{2}+1}\right)\right)^2 & \mbox{ for } & j\geq 2\mbox{ even, }\\[10pt] 0 & \mbox{ for } & j\geq 2\mbox{ odd; } \end{matrix}\right.\\[5pt]
\hat{\eta}_j&=\left\{\begin{matrix} \dfrac{1}{j+3}\left(\dfrac{d}{dx}\left(\tilde{\rho}_{\frac{j}{2}+1}x^{\frac{j}{2}+1}\right)\right)^2 & \quad\,\,\mbox{ for } & j\geq 0\mbox{ even, }\\[10pt] \,\,\,\,\,\,\,0 & \quad\,\,\mbox{ for } & j\geq 0\mbox{ odd; } \end{matrix}\right.\\
\end{split}
\end{equation}
\begin{equation}\label{quant_j}
\begin{split}
\hat{\zeta}_j&=\frac{j(j+2)}{4(j+1)}\tilde{\rho}^2_\frac{j}{2}x^j\\
&-\slim_{k=2}^{j-2}\slim_{m=1}^{k-1}\slim_{\alpha=0}^{\left\lfloor\frac{m}{2}\right\rfloor}\slim_{\beta=0}^{\left\lfloor\frac{k-m}{2}\right\rfloor}C^{m,k}_{\alpha,\beta}\tilde{\rho}_m\tilde{\rho}_{k-m}\tilde{\sigma}_{j-k}x^j\int\limits_{-1}^{1}P^0_{m-1-2\alpha}P^0_{k-m-1-2\beta}P^1_{j-k-1}P^1_1d\xi\\
&+\frac{1}{2}\slim_{k=5}^{j-1}\slim_{m=4}^{k-1}\slim_{p=2}^{m-2}\slim_{\alpha=0}^{\left\lfloor\frac{p}{2}\right\rfloor}\slim_{\beta=0}^{\left\lfloor\frac{m-p}{2}\right\rfloor}C^{p,m}_{\alpha,\beta}\tilde{\rho}_p\tilde{\rho}_{m-p}\tilde{\sigma}_{k-m}\tilde{\sigma}_{j-k}x^j\int\limits_{-1}^{1}P^0_{p-1-2\alpha}P^0_{m-p-1-2\beta}P^1_{k-m-1}P^1_{j-k-1}d\xi;\\
\hat{P}_j&=\frac{1}{2}\slim_{k=2}^{j-2}\tilde{\sigma}_k\tilde{\sigma}_{j-k}x^j\int\limits_{-1}^1\od{}{\xi}\left(P^1_1P^1_{k-1}\right)\od{}{\xi}\left(P^1_1P^1_{j-k-1}\right)d\xi\\
&-\frac{1}{2}\slim_{k=4}^{j-2}\slim_{m=2}^{k-2}\tilde{\sigma}_m\tilde{\sigma}_{k-m}\tilde{\sigma}_{j-k}x^j\int\limits_{-1}^1\od{}{\xi}\left(P^1_1P^1_{m-1}\right)\od{}{\xi}\left(P^1_{k-m-1}P^1_{j-k-1}\right)d\xi\\
&+\frac{1}{8}\slim_{k=6}^{j-2}\slim_{m=4}^{k-2}\slim_{p=2}^{m-2}\tilde{\sigma}_p\tilde{\sigma}_{m-p}\tilde{\sigma}_{k-m}\tilde{\sigma}_{j-k}x^j\int\limits_{-1}^1\od{}{\xi}\left(P^1_{p-1}P^1_{m-p-1}\right)\od{}{\xi}\left(P^1_{k-m-1}P^1_{j-k-1}\right)d\xi;
\end{split}
\end{equation}
defining the constants
\begin{equation}
C^{\,a,b}_{\alpha,\beta}\equiv\left(2(a-1-2\alpha)+1\right)\left(2(b-a-1-2\beta)+1\right).
\end{equation}
We have simplified $\eta$ and $G$ using orthogonality properties of Legendre functions. For $\zeta$ we use the relation
\begin{equation}\label{orth}
\od{P^0_{n+1}}{\xi}=\slim_{\alpha=0}^{\left\lfloor\frac{n+1}{2}\right\rfloor}\left(2(n-2\alpha)+1\right)P^0_{n-2\alpha};
\end{equation}
and for $\zeta$ and $P$, we use the fact that $P^1_1=-\sqrt{1-\xi^2}$. We may also use properties of Legendre functions to expand the $\xi$-derivatives in $\hat{P}_j$, leaving other Legendre functions; and we can write the integrals in $\zeta$ and $P$ explicitly in terms of Wigner's $3-j$-symbol, using results from \cite{dong_overlap_2002} concerning the integral over $\xi\in\mc{I}$ of arbitrary-length products of Legendre functions of arbitrary order and degree. The main point is that these integrals w.r.t. $\xi$ will converge over the finite range $\xi\in\mc{I}$.

We perform a similar analysis of Equations \eqref{RRm} to \eqref{quant_j} as we did in Section \ref{BCs}, where we examined the terms appearing on the right-hand sides, and we may conclude that these equations are consistent in the same sense. We know the gauge field to all orders, and so the set of expansion functions $\{\hat{m}_j(x),\hat{S}_j(x)\}$ may be recursively calculated from the gauge field expansion coefficients and from the set of functions $\{\hat{m}_k(x),\hat{S}_k(x)\}$ for $k<j$. Therefore we may in principle solve equations \eqref{RRm}, \eqref{RRS} sequentially, provided the integral w.r.t. $x$ exists, yielding globally regular solutions -- we return to this point in Section \ref{glomet}.

\subsection{Convergence of the gauge sector}\label{converge}

%
As can be seen from Subsection \ref{LL}, we have discovered solutions which can be expressed as an asymptotic expansion in terms of $\ell$, and that we can calculate the gauge field exactly to all orders. However there is a further complication. Each new order of the expansion introduces one extra constant for each gauge field, and we know that in order for this solution to be truly non-trivial, we need all degrees of freedom of the solution to be represented, which means we shall need \emph{all} constants to be taken into account. This means that we are unable to terminate the gauge field expansions in \eqref{dyonLLexp} at any finite value of $k$ if we want a genuine $\suinf$ solution. 

Therefore, what we now wish to prove is that the gauge field sums in \eqref{dyonLLexp} converge for all values of $x\in\mc{R}$ and $\xi\in\mc{I}$ \eqref{RI} when we take all expansion terms into account. To this end, we prove a series of lemmata.

\begin{lem}\label{P/P}
	For all $\xi\in\mc{I}$,
	\begin{equation}
	\lim_{k\rightarrow\infty}\Bigg|\frac{P^1_{k}(\xi)}{P^1_{k-1}(\xi)}\Bigg|=	\lim_{k\rightarrow\infty}\Bigg|\frac{P^0_{k+1}(\xi)}{P^0_{k}(\xi)}\Bigg|=1.
	\end{equation}
\end{lem}
\textbf{Proof} For fixed $m\geq0$ and $k\rar\infty$, we have the following asymptotic approximation (see 14.15.11 in \cite{olver_nist_2010}) in terms of Bessel functions $J_m(\xi)$, which we have rewritten slightly:
\begin{equation}
P^m_k(\cos\vartheta)=\frac{(-1)^m(k+m)!}{k^{m}(k-m)!}\left(\frac{\vartheta}{\sin\vartheta}\right)^{\frac{1}{2}}J_m\left(\left(k+\frac{1}{2}\right)\vartheta\right)\left[1+O\left(k^{-1}\right)\right],
\end{equation}
for $\vartheta\in[0,\pi)$. Substituting in a definition from Abramowitz and Stegun \cite{abramowitz_handbook_1988} (9.1.69), again rewritten slightly,
\begin{equation}
J_m(z)=\frac{2^{m+2}z^m}{\Gamma(m+1)\left(4+z^2\right)^{m+1}},
\end{equation}
it is easy enough to see that
\begin{equation}\label{Plims1}
\begin{split}
\lim_{k\rightarrow\infty}&\Bigg|\frac{P^1_{k}(\xi)}{P^1_{k-1}(\xi)}\Bigg|=\lim_{k\rightarrow\infty}\Bigg|\left(\frac{k+1}{k}\right)\left(\frac{k+\frac{1}{2}}{k-\frac{1}{2}}\right)\left(\frac{\left(\left(k-\frac{1}{2}\right)\vartheta\right)^2+4}{\left(\left(k+\frac{1}{2}\right)\vartheta\right)^2+4}\right)^2\left(\frac{1+O\left(k^{-1}\right)}{1+O\left(k^{-1}\right)}\right)\Bigg|\\[5pt]
&=\lim_{k\rightarrow\infty}\Bigg|\left(1+k^{-1}\right)\left(\frac{1+\frac{k^{-1}}{2}}{1-\frac{k^{-1}}{2}}\right)\left(\frac{\left(\left(1-\frac{k^{-1}}{2}\right)\vartheta\right)^2+4k^{-2}}{\left(\left(1+\frac{k^{-1}}{2}\right)\vartheta\right)^2+4k^{-2}}\right)^2\left(\frac{1+O\left(k^{-1}\right)}{1+O\left(k^{-1}\right)}\right)\Bigg|\\
&=1,\quad\forall\,\xi\in\mc{I}.
\end{split}
\end{equation}
The proof for the Legendre polynomials $P^0_k$ is similar but simpler. $\Box$

As for the functions of $x$, we can easily prove the following results using the properties of hypergeometric functions: 
\begin{equation}\label{siglim}
\begin{split}
&\lim_{x\rar 0}\frac{x\mc{S}_{k+1}}{\mc{S}_k}=\lim_{x\rar 0}\frac{x\mc{P}_{k+1}}{\mc{P}_k}=0\quad\forall\,k\geq 2,\\[5pt]
\lim_{k\rar\infty}&\lim_{x\rar \infty}\frac{x\mc{S}_{k+1}}{\mc{S}_k}=\lim_{k\rar\infty}\lim_{x\rar \infty}\frac{x\mc{P}_{k+1}}{\mc{P}_k}=2.
\end{split}
\end{equation}
In addition, some numerical investigation seems to indicate, for large $k$ (and indeed all $k\geq 2$ surveyed) and throughout the entire range of $x$, that the ratios $\frac{x\mc{S}_{k+1}}{\mc{S}_k}$ and $\frac{x\mc{P}_{k+1}}{\mc{P}_k}$ are bounded above and below by these boundary values. We now demonstrate this in the following two Lemmata.

\begin{lem}\label{sig/sig}
	For all $x\in\mc{R}$,
	\begin{equation}\label{Sratio}
	0\leq\lim_{k\rar\infty}\frac{x\mc{S}_{k+1}(x)}{\mc{S}_k(x)}\leq 2.
	\end{equation}
\end{lem}

\textbf{Proof} Firstly, we know that if we have $a,b,c>0$, then $0<\,\!_2F_1(a,b;c;z)\leq1$ for $z\in(-\infty,0]$, which implies that $0<\mc{S}_k<1$ for all $k$ and all $x\in\mc{R}$, so it is easy to prove that the lower bound is respected. To prove the condition obeys the upper bound, we rewrite \eqref{Sratio} as
\begin{equation}\label{Sratio1}
\lim_{k\rar\infty}\left[2\mc{S}_k-x\mc{S}_{k+1}\right]\geq 0,\,\,\forall\,x\in\mc{R}.
\end{equation}
Using \eqref{2F1asym}, and rearranging some gamma functions, we may write 
\begin{equation}\label{Sratio2}
\begin{split}
2\mc{S}_k-x\mc{S}_{k+1}&=\sqrt{\pi}x^{-k-1}\frac{\Gamma\left(\frac{2k+1}{2}\right)}{\Gamma\left(\frac{k+1}{2}\right)^2}\left\{x\left[2F^{\frac{k}{2},\frac{1-k}{2}}_{\frac{1}{2}}-\frac{2k+1}{2}\left(\frac{\Gamma\left(\frac{k+1}{2}\right)}{\Gamma\left(\frac{k+2}{2}\right)}\right)^2F^{\frac{k+1}{2},\frac{-k}{2}}_{\frac{1}{2}}\right]\right.\\
&\left.+\,2\left[\frac{2k+1}{2}F^{\frac{k+2}{2},\frac{1-k}{2}}_{\frac{3}{2}}-2\left(\frac{\Gamma\left(\frac{k+1}{2}\right)}{\Gamma\left(\frac{k}{2}\right)}\right)^2F^{\frac{k+1}{2},\frac{2-k}{2}}_{\frac{3}{2}}\right]\right\},
\end{split}
\end{equation}
where for reasons of space we are using the shorthand
\begin{equation}
F^{a,b}_c(x)\equiv\,_2F_1\left(a,b;c;-x^{-2}\right).
\end{equation}

Our strategy is as follows. We fix $x=x_0$ to be some (finite, non-zero) value, and we use the definition of hypergeometric functions in terms of gamma functions, which in our case is
\begin{equation}\label{FPoch}
F^{a,b}_c(x_0)=\frac{\Gamma(c)}{\Gamma(a)\Gamma(b)}\slim_{n=0}^\infty\frac{\Gamma(a+n)\Gamma(b+n)}{\Gamma(c+n)}\frac{(-x_0^{-2})^n}{\Gamma(n+1)},
\end{equation}
to write \eqref{Sratio2} out as one large summation, considering it as a function parametrised by $k$. Then we will consider a general term in the summation. We will prove that for $k$ large, \emph{each} term of the summation is positive, implying that so is the total sum. In anticipation of taking the limit $k\rar\infty$, we also consider that for the summation term indexed by $n$ that we will consider, we have $n<<k$; this ensures that none of the sums in \eqref{FPoch} terminate, simplifying the situation.

Rewriting \eqref{Sratio2} in accordance with the above recipe, we arrive at the following condition for each term of the total summation of the hypergeometric functions: For each $n\in\mathbb{N}$, for every $x_0\in\mc{R}$ and in the limit $k\rar\infty$, we wish that
\begin{equation}\label{Sratio3}
\begin{split}
&\sqrt{\pi}x_0^{-k-2n-1}\frac{\Gamma\left(\frac{2k+1}{2}\right)}{\Gamma\left(\frac{k+1}{2}\right)^2}(-1)^n\left\{	x_0\left(\frac{\Gamma\left(\frac{1}{2}\right)}{\Gamma\left(n+\frac{1}{2}\right)\Gamma(n+1)}\right)\left[\frac{2\Gamma\left(\frac{k}{2}+n\right)}{\Gamma\left(\frac{k}{2}\right)}\frac{\Gamma\left(\frac{1-k}{2}+n\right)}{\Gamma\left(\frac{1-k}{2}\right)}\right.\right.\\
&\left.\left.-\frac{2k+1}{2}\left(\frac{\Gamma\left(\frac{k+1}{2}\right)}{\Gamma\left(\frac{k+2}{2}\right)}\right)^2\frac{\Gamma\left(\frac{k+1}{2}+n\right)}{\Gamma\left(\frac{k+1}{2}\right)}\frac{\Gamma\left(\frac{-k}{2}+n\right)}{\Gamma\left(\frac{-k}{2}\right)}\right]+2\left(\frac{\Gamma\left(\frac{3}{2}\right)}{\Gamma\left(n+1\right)\Gamma(n+\frac{3}{2})}\right)\right.\\
&\left.\times\left[\frac{2k+1}{2}\frac{\Gamma\left(\frac{k+2}{2}+n\right)}{\Gamma\left(\frac{k+2}{2}\right)}\frac{\Gamma\left(\frac{1-k}{2}+n\right)}{\Gamma\left(\frac{1-k}{2}\right)}-2\left(\frac{\Gamma\left(\frac{k+1}{2}\right)}{\Gamma\left(\frac{k}{2}\right)}\right)^2\frac{\Gamma\left(\frac{k+1}{2}+n\right)}{\Gamma\left(\frac{k+1}{2}\right)}\frac{\Gamma\left(\frac{2-k}{2}+n\right)}{\Gamma\left(\frac{2-k}{2}\right)}\right]\right\}\geq 0.
\end{split}
\end{equation}
Note that we can simplify two of the terms in the parentheses -- the numerators by using $\Gamma(\frac{3}{2})=\frac{1}{2}\Gamma(\frac{1}{2})=\frac{\sqrt{\pi}}{2}$, and their respective denominators with the so-called duplication formula $\Gamma(z)\Gamma(z+\frac{1}{2})=2^{1-2z}\sqrt{\pi}\,\Gamma(2z)$:
\begin{equation}\label{dupe}
\frac{\Gamma\left(\frac{1}{2}\right)}{\Gamma\left(n+\frac{1}{2}\right)\Gamma(n+1)}=\frac{2^{2n}}{\Gamma(2n+1)},\qquad\frac{\Gamma\left(\frac{3}{2}\right)}{\Gamma(n+1)\Gamma\left(n+\frac{3}{2}\right)}=\frac{2^{2n}}{\Gamma(2n+2)}.
\end{equation}

Our aim is now to use the asymptotic formula
\begin{equation}\label{gammasym}
\lim_{a\rar\infty}\frac{\Gamma(a+b)}{\Gamma(a)a^b}=1,
\end{equation}
by `pulling out' factors from each square-bracketed term in \eqref{Sratio3}, which all involve quotients of gamma functions, such that we can apply \eqref{gammasym} in the limit $k\rar\infty$. We shall demonstrate this explicitly for the second term in the first set of square brackets, 
\begin{equation}\label{2ndterm}
-(-1)^n\frac{2k+1}{2}\left(\frac{\Gamma\left(\frac{k+1}{2}\right)}{\Gamma\left(\frac{k+2}{2}\right)}\right)^2\frac{\Gamma\left(\frac{k+1}{2}+n\right)}{\Gamma\left(\frac{k+1}{2}\right)}\frac{\Gamma\left(\frac{-k}{2}+n\right)}{\Gamma\left(\frac{-k}{2}\right)},
\end{equation}
since the others are similar and simpler. Note that we have deliberately included the term $(-1)^n$ to eliminate unwanted negative signs. The final factor in \eqref{2ndterm} is problematic, because it contains gamma functions with arguments that tend to $-\infty$ as $k\rar\infty$. To rewrite it, we make use of an identity of gamma functions,
\begin{equation}
\frac{\Gamma(-m+n)}{\Gamma(-m)}\equiv\frac{(-1)^{n}m\Gamma(m)}{(m-n)\Gamma(m-n)},
\end{equation}
derived from Euler's reflection formula. Letting $m=\frac{k}{2}$ and rearranging, we find
\begin{equation}
\frac{\Gamma\left(\frac{-k}{2}+n\right)}{\Gamma\left(\frac{-k}{2}\right)}=\frac{(-1)^{n}(\frac{k}{2})\Gamma(\frac{k}{2})}{\left(\frac{k}{2}-n\right)\Gamma\left(\frac{k}{2}-n\right)}.
\end{equation}
Substituting this into \eqref{2ndterm}, and by `pulling out' factors, we obtain
\begin{equation}\label{2ndterm1}
\begin{split}
&-\frac{2k+1}{2}\frac{2}{k+1}\left(\frac{k+1}{2}\right)^n\left(\frac{k}{2}\right)^n\frac{\frac{k}{2}}{\frac{k}{2}-n}\\
&\times\left[\frac{\Gamma\left(\frac{k+1}{2}\right)\left(\frac{k+1}{2}\right)^\frac{1}{2}}{\Gamma\left(\frac{k+2}{2}\right)}\right]^2\left[\frac{\Gamma\left(\frac{k+1}{2}+n\right)}{\Gamma\left(\frac{k+1}{2}\right)\left(\frac{k+1}{2}\right)^n}\right]\left[\frac{\Gamma\left(\frac{k}{2}\right)\left(\frac{k}{2}\right)^{-n}}{\Gamma\left(\frac{k}{2}-n\right)}\right].
\end{split}
\end{equation}
Note that using \eqref{gammasym}, all the terms in \eqref{2ndterm1} in square brackets will equal 1 in the limit $k\rar\infty$. Continuing in this vein, Expression \eqref{2ndterm1} becomes
\begin{equation}\label{2ndterm2}
\begin{split}
&\,-\left(\frac{k^{2n}}{2^{2n}}\right)\left(\frac{2+k^{-1}}{1+k^{-1}}\right)\frac{(1+k^{-1})^n}{1-2nk^{-1}}\\
&\times\left[\frac{\Gamma\left(\frac{k+1}{2}\right)\left(\frac{k+1}{2}\right)^\frac{1}{2}}{\Gamma\left(\frac{k+2}{2}\right)}\right]^2\left[\frac{\Gamma\left(\frac{k+1}{2}+n\right)}{\Gamma\left(\frac{k+1}{2}\right)\left(\frac{k+1}{2}\right)^n}\right]\left[\frac{\Gamma\left(\frac{k}{2}\right)\left(\frac{k}{2}\right)^{-n}}{\Gamma\left(\frac{k}{2}-n\right)}\right].
\end{split}
\end{equation}
We perform a similar procedure on the other square-bracketed terms in \eqref{Sratio3}, and see that a factor of $k^{2n}$ may be taken out of each of them, as in \eqref{2ndterm2}. Then we distribute the limit over some of the factors in \eqref{Sratio3}, because this will clarify the situation. In the limit $k\rar\infty$, the first square bracket in Equation \eqref{Sratio3} vanishes, and the second becomes equal to $\frac{1}{2}$; and therefore all the terms in the braces simplify completely to $\frac{1}{\Gamma(2n+2)}$. Therefore, we obtain the condition
\begin{equation}\label{finalc}
\left(\frac{\sqrt{\pi}}{\Gamma(2n+2)}\right)\lim_{k\rar\infty}\left[\frac{k^{2n}\Gamma\left(\frac{2k+1}{2}\right)}{x_0^{k+2n+1}\Gamma\left(\frac{k+1}{2}\right)^2}\right]\geq 0.
\end{equation}
We might be able to simplify this further, but we can already see that all the terms on the left-hand side of \eqref{finalc} are positive; meaning that \eqref{Sratio1} must be satisfied for all $n$, even as $n\rar\infty$, and for all $x_0\in(0,\infty)$. Therefore the Lemma is proven. 
$\Box$

\begin{lem}\label{rho/rho}
	For all $x\in\mc{R}$,
	\begin{equation}
	0\leq\lim_{k\rar\infty}\Bigg|\frac{x\mc{P}_{k+1}(x)}{\mc{P}_k(x)}\Bigg|\leq 2.
	\end{equation}
\end{lem}
The proof of this Lemma is extremely similar to that for Lemma \ref{sig/sig}. The lower limit is elementary; and for the upper limit, we end up with a condition similar to \eqref{finalc},
\begin{equation}\label{finalc2}
\left(\frac{\sqrt{\pi}}{\Gamma(2n+2)}\right)\lim_{k\rar\infty}\left[\frac{k^{2n}\Gamma\left(\frac{2k+3}{2}\right)}{x_0^{k+2n+1}\Gamma\left(\frac{k+1}{2}\right)\Gamma\left(\frac{k+3}{2}\right)}\right]\geq 0,
\end{equation}
which is again satisfied for all $n$ (including as $n\rar\infty$) and all $x_0\in(0,\infty)$.

Finally, we can now easily prove the main result of this Subsection:
\begin{prop}\label{conv}
	The infinite sums defining the gauge functions $\alpha(r,\xi)$ and $\omega(r,\xi)$ \eqref{dyonLLexp} converge, provided that the following conditions on the constants $\{\rho_k,\sigma_k\}$ are satisfied:
	
	\begin{equation}\label{concon}
	\lim_{k\rightarrow\infty}\Bigg|\frac{\sigma_{k+1}}{\sigma_k}\Bigg|<\frac{1}{2\ell},\qquad\lim_{k\rightarrow\infty}\Bigg|\frac{\rho_{k+1}}{\rho_k}\Bigg|<\frac{1}{2\ell}.
	\end{equation}
\end{prop}

\textbf{Proof} We use the ratio test to assess the convergence of the sums in \eqref{dyonLLexp}. The ratio test for the convergence of $\omega(r,\xi)$ gives the following condition for all $x\in\mc{R}$, $\xi\in\mc{I}$:
\begin{equation}\label{magcon}
\lim_{k\rightarrow\infty}\Bigg|\frac{\sigma_{k+1}\mc{S}_{k+1}(x)P^1_{k}(\xi)(x\ell)^{k+1}}{\sigma_k\mc{S}_k(x)P^1_{k-1}(\xi)(x\ell)^{k}}\Bigg|=\lim_{k\rightarrow\infty}\Bigg|\ell\frac{\sigma_{k+1}}{\sigma_k}\Bigg|\Bigg|\frac{x\mc{S}_{k+1}}{\mc{S}_k}\Bigg|\Bigg|\frac{P^1_{k}}{P^1_{k-1}}\Bigg|
<1,
\end{equation} 
and using Lemmata \ref{P/P} and \ref{sig/sig}, this quickly becomes the result stated in the theorem. The proof for the condition on the electric field $\alpha(r,\xi)$ is very similar to prove starting with
\begin{equation}\label{elecon}
\lim_{k\rar\infty}\Bigg|\frac{\rho_{k+1}\mc{P}_{k+1}(x)P^0_{k+1}(\xi)(x\ell)^{k+1}}{\rho_k\mc{P}_k(x)P^0_{k}(\xi)(x\ell)^k}\Bigg|<1\,\,\forall x\in\mc{R},\,\,\xi\in\mc{I},
\end{equation}
and using Lemmata \ref{P/P} and \ref{rho/rho}. $\Box$

It is pleasing to remark that Proposition \ref{conv} gives enormous latitude to the sets of constants $\{\rho_k\}$ and $\{\sigma_k\}$ that will satisfy the convergence conditions. Since we are interested in small $\ell$, the conditions will be easy to satisfy even if both series of constants are geometrically increasing, as long as the geometric ratio is less than $\frac{1}{2\ell}$ as $k\rar\infty$. In fact in the limit as $\ell\rar0$, \emph{any} choice of constants will satisfy convergence. Again, this is familiar from the $\sun$ case \cite{baxter_soliton_2007}.

\subsection{Global regularity of the metric sector}\label{glomet}

Due to the sheer complexity, there is a lot less we can do with the metric functions $m$ and $S$. It is a task beyond the scope of this paper to prove that the metric expansions will converge to all orders in $\ell$. However, it is not as important to us to have exact solutions for $m$ and $S$, knowing that the infinite set of gauge degrees of freedom are already represented in the solution. Hence, we have the freedom to consider an approximate solution for the metric functions with an arbitrarily large number of terms and hence to a high degree of accuracy, as long as we can show the Einstein equations may be solved to give functions which are globally regular. That is our aim here.

\begin{prop}
	The recursive differential equations \eqref{RRm}, \eqref{RRS} may be sequentially solved by integration and substitution, to arbitrarily high order, using the lowest order metric equations \eqref{Ein2} as a starting point. The resulting functions $\hat{m}_j(x)$ and $S_j(x)$ are globally regular for all $j\geq 2$.
\end{prop}

\textbf{Proof} We note that have explicit expressions for the gauge fields to all expansion orders (\ref{r0LLrho}, \ref{r0LLsig}); and that the equations \eqref{RRm} and \eqref{RRS} are consistent, in that if we possess knowledge of the metric functions up to order $j$, we may write explicitly the equations for order $j+1$. Our task now will essentially be to show that the equations for the metric functions may be solved by integration at every order, to give globally regular solutions. The integrals of \eqref{RRm} and \eqref{RRS} w.r.t $x$ will exist for all $x\in\mc{R}$ provided that (for all $j$),
\begin{equation}\label{intsplit}
\od{\hat{m}_j}{x}\sim O(x^p)\mbox{ for }p\geq 0\mbox{ when }x\sim 0\quad\mbox{and}\quad\od{\hat{m}_j}{x}\sim O(x^{-p})\mbox{ for }p\geq 2\mbox{ as }x\rar\infty,
\end{equation}
and similar for $\od{\hat{S}_j}{x}$. We will now show that this is the case.

To do this, we use an inductive argument. Given that we obtained the boundary conditions for the equations for $\hat{m}_2(x)$ and $\hat{S}_2(x)$, we make the inductive hypothesis that given some integer $J$, then for all integers $2<j\leq J$, then $\hat{m}_j(x)$ and $\hat{S}_j(x)$ are globally regular; and that we have the following behaviour at the spatial boundaries, motivated by the results of Section \ref{BCs}. When $x\sim 0$, we assume that we have
\begin{equation}\label{ind0}
\begin{split}
\od{\hat{m}_j}{x}\sim O(x^2)\quad&\implies\hat{m}_j(x)\sim O(x^3)
,\\
\od{\hat{S}_j}{x}\sim O(x)\quad\,\,\,\,&\implies \hat{S}_j(x)\sim S_{j,0}+O(x^2)
;
\end{split}
\end{equation}
and as $x\rar\infty$,
\begin{equation}\label{indinf}
\begin{split}
\od{\hat{m}_j}{x}\sim O(x^{-2})\quad&\implies\hat{m}_j(x)\sim M_j+O(x^{-1}),\\
\od{\hat{S}_j}{x}\sim O(x^{-5})\quad\,&\implies \hat{S}_j(x)\sim O(x^{-4});
\end{split}
\end{equation}
using the conditions \eqref{mkSkBC}. Noting \eqref{elltrans}, the integration constants $S_{j,0}\leq0$ and $M_j\geq0$, when known to all orders, fix the remaining metric function boundary behaviour not fixed by \eqref{mkSkBC}:
\begin{equation}\label{M}
M=\sum_{j=2}^\infty M_j\ell^{j+1}\sim O(\ell^3),\qquad S_0=1+\sum_{j=2}^\infty S_{j,0}\ell^{j}\sim 1+O(\ell^2).
\end{equation}
Now we consider the differential recurrence equations \eqref{RRm} and \eqref{RRS}. Using the inductive hypothesis, we may show that in \eqref{widehat}, we have the following behaviour: For $x\sim 0$,
\begin{equation}\label{M_0lim}
\widehat{\mc{M}}_j(x)\sim O(1),\qquad\widetilde{\mc{M}}_j(x)\sim O(1);
\end{equation}
and as $x\rar\infty$,
\begin{equation}\label{M_inflim}
\widehat{\mc{M}}_j(x)\sim O(x^{-1}),\qquad\widetilde{\mc{M}}_j(x)\sim O(1),
\end{equation}
for all $j\leq J$.

In addition, we may use the boundary behaviour of the gauge field expansions \eqref{GFbounds} to calculate the behaviour of the functions $\eta_j$, $G_j$, $\zeta_j$ and $P_j$ \eqref{quant_j} near the spatial boundaries: When $x\sim 0$, we obtain
\begin{equation}\label{quants_0lim}
\begin{matrix}
G_j\sim O(x^{j})\,\,\,(2\leq j\leq J), & \quad & \eta_j\sim O(x^{j})\,\,\,(0\leq j \leq J),\\[7pt]
\zeta_j\sim O(x^j)\,\,\,(2\leq j \leq J), & \quad & P_j\sim O(x^{j})\,\,\,(4\leq j \leq J);\\
\end{matrix}
\end{equation}
%
and as $x\rar\infty$,
\begin{equation}\label{quantsinflim}
G_j\sim O(x^{-4}),\quad\eta_j\sim O(x^{-4}),\quad\zeta_j\sim O(1),\quad P_j\sim O(1).
\end{equation}

Finally, examining the recursion equations \eqref{RRm} and \eqref{RRS} in light of (\ref{M_0lim}--\ref{quantsinflim}), we can determine that
\begin{equation}
\od{\hat{m}_{J+1}}{x}\sim O(x^2),\quad
\od{\hat{S}_{J+1}}{x}\sim O(x);
\end{equation}
and as $x\rar\infty$,
\begin{equation}
\od{\hat{m}_{J+1}}{x}\sim O(x^{-2}),\quad
\od{\hat{S}_{J+1}}{x}\sim O(x^{-5}).
\end{equation}
Examining \eqref{ind0} and \eqref{indinf}, this completes our induction step. Therefore, with the boundary conditions for $\od{\hat{m}_j}{x}$ and $\od{\hat{S}_j}{x}$ proven for all $j\geq 2$, we can integrate \eqref{RRm} and \eqref{RRS} with respect to $x$ to obtain globally regular solutions to all expansion orders, and the Proposition is proven. $\Box$

Finally, we can collect the results of this Section into the following
\begin{thr}
	Solutions exist to the field equations (\ref{EEs}, \ref{YMs}) which are globally regular, i.e. throughout the range $(r,\xi)\in\mc{R}\times\mc{I}$, in the limit as $|\Lambda|\rar\infty$. All field variables possess the correct boundary behaviour as outlined in Section \ref{BCs}. 
	
	The gauge field functions $\alpha(r,\xi)$ and $\omega(r,\xi)$ may be explicitly written to all orders (\ref{r0LLrho}, \ref{r0LLsig}), assuming we specify the constants $\{\rho_k,\sigma_k\}$ and the value of $\ell$. These constants comprise the infinite number of degrees of freedom of the gauge field variables, one appearing at each order of expansion in $\ell$. Moreover, the infinite expansions defining the gauge fields \eqref{dyonLLexp} converge. 
	
	The metric functions $m(r)$ and $S(r)$ are globally regular at each order in $\ell$ and may be calculated recursively to arbitrarily high order; and therefore if the expansion sums of the metric functions do not converge, we may at least take a truncated approximation to arbitrarily high order in $\ell$. If they do converge, then we have an exact solution to the field equations.
\end{thr}

\section{Characterising solutions uniquely by effective global charges}\label{Charges}

In the case of black hole solutions to the field equations \eqref{EEs}, \eqref{YMs}, we were able to define effective charges which are calculated from the asymptotic boundary functions, and uniquely characterised the solutions we found in the limit $|\Lambda|\rar\infty$. The same argument for the derivation of the charge expressions carries over in this case. Therefore, we again have the charge functions
\begin{equation}\label{chargebij}
q_E(\xi)=\frac{\sqrt{3}}{2}\mc{A}(\xi),\qquad q_M(\xi)=\frac{\sqrt{3}}{2}\od{}{\xi}\left(\omega_\infty(\xi)^2+\xi^2\right),
\end{equation}
and total squared charges
\begin{equation}\label{charges}
\mc{Q}^2_E=\frac{1}{2}\int\limits_{-1}^1q_E^2 d\xi,\qquad
\mc{Q}^2_M=\frac{1}{2}\int\limits_{-1}^1q_M^2d\xi.
\end{equation}
We proved in \cite{baxter_inprep} that the functions \eqref{chargebij} uniquely define the asymptotic boundary data functions $\{\omega_\infty,\mc{A}\}$ defined by \eqref{infexp}.

Substituting \eqref{r0LLrho} and \eqref{r0LLsig} into \eqref{dyonLLexp}, and using \eqref{2F1asym}, the asymptotic boundary data for the gauge fields can be calculated as follows:
\begin{equation}\label{asymgauge}
\begin{split}
\alpha_\infty(\xi)&=\sqrt{\pi}\slim_{k=1}^\infty\frac{\rho_k\,\Gamma\left(\frac{2k+3}{2}\right)}{\Gamma\left(\frac{k+1}{2}\right)\Gamma\left(\frac{k+3}{2}\right)}P^0_k(\xi)\ell^k,\\
\mc{A}(\xi)&=2\sqrt{\pi}\slim_{k=2}^\infty\rho_{k-1}(k-1)\left(\frac{\Gamma\left(\frac{2k+1}{2}\right)}{\Gamma\left(\frac{k-1}{2}\right)\Gamma\left(\frac{k+1}{2}\right)}-\frac{k\Gamma\left(\frac{2k+3}{2}\right)}{(2k+1)\Gamma\left(\frac{k+1}{2}\right)^2}\right)P^0_{k-1}(\xi)\ell^k,\\
\omega_\infty(\xi)&=\sqrt{1-\xi^2}+\sqrt{\pi}\slim_{k=2}^\infty\frac{\sigma_k\,\Gamma\left(\frac{2k+1}{2}\right)}{\Gamma\left(\frac{k+1}{2}\right)^2}P^1_{k-1}(\xi)\ell^k,\\
\mc{W}(\xi)&=2\sqrt{\pi}\slim_{k=3}^\infty\sigma_{k-1}(k-1)\left(\frac{ \Gamma\left(\frac{2k-1}{2}\right)}{\Gamma\left(\frac{k-1}{2}\right)^2}-\frac{k\Gamma\left(\frac{2k+1}{2}\right)}{(2k-1)\Gamma\left(\frac{k-1}{2}\right)\Gamma\left(\frac{k+1}{2}\right)}\right)P^1_{k-2}(\xi)\ell^k,
\end{split}
\end{equation}
where $\{\rho_k,\sigma_k\}$ are the constants which define the solution near $x=0$. Since the boundary functions are all sums over bases of orthogonal polynomials, we immediately see that these functions are uniquely fixed by the choice of constants $\{\rho_k,\sigma_k\}$. In fact, we may use the orthogonality of Legendre functions \eqref{orth} to invert two of these relationships:
\begin{equation}
\begin{split}
\rho_k=\,\, &\frac{2k+1}{4\sqrt{\pi}k\ell^{k+1}}\left(\frac{(2k+3)\Gamma\left(\frac{k}{2}\right)\Gamma\left(\frac{k+2}{2}\right)^2}{(k+1)\Gamma\left(\frac{k}{2}\right)\Gamma\left(\frac{2k+5}{2}\right)-(2k+3)\Gamma\left(\frac{2k+3}{2}\right)\Gamma\left(\frac{k+2}{2}\right)}\right)\times\\
&\int\limits_{-1}^1\mc{A}(\xi)P^0_k(\xi)d\xi\,\,\,\mbox{    for }k\geq 1,\\
\sigma_k&=\frac{(2k-1)\Gamma\left(\frac{k+1}{2}\right)^2}{2\sqrt{\pi}k(k-1)\ell^k\Gamma\left(\frac{2k+1}{2}\right)}\int\limits_{-1}^1\omega_\infty(\xi)P^1_{k-1}(\xi)d\xi+\frac{1}{3\ell^2}\delta_{2,k}\,\,\,\mbox{    for }k\geq 2.
\end{split}
\end{equation}
Thus it can be shown that knowledge of the asymptotic boundary functions $\{\mc{A}(\xi),\omega_\infty(\xi)\}$ uniquely fixes the constants $\{\rho_k,\sigma_k\}$ and hence entirely specifies the solution at both boundaries. Therefore, so do the charge functions \eqref{chargebij}.

It is important to note that just as in the black hole case, the charges \eqref{charges} act as \emph{effective charges} in the limit $\ell\rar 0$, in the sense that the total squared charge $\mc{Q}^2=\mc{Q}_{M}^2+\mc{Q}^2_E$ will coincide with the quantity $Q_{eff}^2$ defined by the asymptotic behaviour of $m$,
\begin{equation}\label{masym}
m(r)=M-\frac{Q^2_{eff}}{2r}+O(r^{-2}),
\end{equation}
i.e. $Q^2_{eff}$ plays the same role in the metric as the Abelian charge. We can see this coincidence as follows. Note that the asymptotic boundary conditions on $m$ (\ref{infexp}, \ref{m1}) imply that $Q_{eff}^2=-2m_1$. Examining \eqref{asymgauge}, we note that $\mc{W}\sim O(\ell^3)$ and $\alpha_\infty\sim O(\ell)$, meaning that the last two terms in $m_1$ are of $O(\ell^4)$. So in the limit $\ell\rar 0$, we have $\mc{Q}^2=-2m_1$, and therefore $\mc{Q}^2=Q_{eff}^2$.

We are also able to asymptotically distinguish black hole solutions from soliton/dyon solutions, by their ADM mass $M$. It can be remarked that when $|\Lambda|\rar\infty$, the magnetic asymptotic data $\omega_\infty(\xi)\sim O(1)$ in the case of both black holes and solitons, so that we can fairly compare the two cases. For black hole solutions, we have
\begin{equation}
M=\frac{r_h^3}{2\ell^2}+\frac{r_h}{2}+\frac{\mc{Q}_M^2}{2r_h}+O(\ell^2),
\end{equation}
i.e. $M\sim O(\ell^{-2})$; but in the soliton/dyon case, considering \eqref{M}, we have $M\sim O(\ell^3)$. Therefore for a fixed small value of $\ell$, globally regular solutions are much lower in mass than corresponding black hole solutions with similar gauge field boundary data. This is something we also find for $\sun$ purely magnetic solutions \cite{shepherd_characterizing_2012}. 
%
We now summarise the results of this Section.
\begin{thr}\label{chargethm}
	We fix the value of $\Lambda<0$ very large, i.e. we fix $\ell\ll1$. Then, globally regular solutions to the full dyonic system are uniquely specified by their charge functions $\{q_M(\xi),q_E(\xi)\}$. In addition, black hole solutions may distinguished from globally regular solutions with similar magnitude gauge field functions by the ADM mass $M$, which is $O(\ell^{-2})$ for black holes and $O(\ell^3)$ for solitons and dyons.
\end{thr}

\section{Conclusions}\label{Conc}

In this work we have proven the existence of globally regular, asymptotically AdS solutions to $\suinf$ EYM field equations \eqref{EEs}, \eqref{YMs}, in the limit as $|\Lambda|\rar\infty$. We also proved that the gauge fields of these solutions are uniquely characterised by an infinite number of arbitrary constants $\{\rho_k,\sigma_k\}$, and that when the gauge functions are expressed as infinite expansions in orders of $\ell$, these sums converge; hence, these solutions genuinely possess an infinite amount of non-trivial hair. In addition, we proved that we obtain a regular Einstein sector at least for some truncation of the series for $m(r)$ and $S(r)$, for which we may calculate an arbitrarily high number of terms. Finally, we established that: i) Dyons (including purely magnetic solitons) are uniquely characterised by their global effective charge functions \eqref{chargebij}; and ii) Black hole solutions and globally regular solutions which possess similar magnitude gauge fields asymptotically may be distinguished from each other by their ADM masses. Our analysis revealed a rich solution structure -- ironically, the complexity of these solutions has in some ways meant that we were able to derive more precise results about them than we obtained for black holes, especially concerning the gauge fields. Our work also suggests, using the analogy that we described between $\suinf$ and $\sun$ in \cite{baxter_inprep}, that slightly better results are available for $\sun$ with $|\Lambda|\rar\infty$ (and therefore, probably for general compact gauge groups) than currently exist \cite{baxter_existence_2016,baxter_existence_2018}: There, we did not obtain the precise form of the electric gauge field, proving only the existence of solutions in the regime unique with respect to the gauge degrees of freedom. This may be worth revisiting.

Future work from this research would most likely involve physical modelling applications -- one possibility is further describing `strange stars' and other exotic gravitational objects, to which we referred in the introduction. On this subject, we return to the conjecture posed in the introduction: Can globally regular $\suinf$ EYM solutions be considered as a model for black hole remnants? This research has only scratched the surface of this question, but has confirmed that these dyons have very small mass and can possess a large number of degrees of freedom, which we expect from remnants. 
To take this project further, we could 
consider coupling a scalar field to this system, possibly also dropping the assumption of staticity, to investigate the evaporation stage of the black hole solutions investigated in \cite{baxter_inprep}, and in light of \cite{hawking_black_1976}. This is a very difficult problem, but it may be of relevance to issues of quantum gravity and the Black Hole Information Paradox.

\vspace{1cm}
\appendix
%
\newpage
\section*{References}
%
%
%


\end{document}